\documentclass[aps,prd,reprint,superscriptaddress,nofootinbib]{revtex4-2}

\usepackage[T1]{fontenc}
\usepackage[utf8]{inputenc}
\usepackage{amsmath,amssymb,mathtools,bm}
\usepackage{graphicx}
\usepackage{xcolor}
\usepackage{tikz}
\usetikzlibrary{arrows.meta,calc,decorations.pathmorphing,positioning,shapes.geometric}
\usepackage[colorlinks=true,linkcolor=blue,citecolor=blue,urlcolor=blue]{hyperref}

\newcommand{\AdS}{\mathrm{AdS}}
\newcommand{\CFT}{\mathrm{CFT}}

\newcommand{\V}{\mathcal V}
\newcommand{\J}{\mathcal J}
\newcommand{\ellthree}{\ell_3}

\newcommand{\Sgen}{S_{\rm gen}}
\newcommand{\dd}{\mathrm d}

\begin{document}

\title{Quantum Black Hole Chemistry from Double Holography}

\author{Naman Kumar}
\affiliation{Department of Physics, Indian Institute of Technology Gandhinagar, Palaj, Gujarat, 382355, India}
\email{namankumar5954@gmail.com, naman.kumar@iitgn.ac.in}

\date{\today}

\begin{abstract}
Extended black hole thermodynamics highlights a sharp tension in the usual holographic dictionary: at fixed boundary conformal frame, changing the AdS radius changes both the central charge and the spatial volume of the CFT, so that the color and volume sectors of the first law becomes degenerate. We show that this degeneracy is naturally removed for quantum black holes in Karch--Randall double holography. The mechanism is intrinsically semiclassical. Replacing the holographic regulator surface by a physical brane induces gravity coupled to a cutoff CFT, so classical bulk black holes become lower-dimensional quantum black holes whose geometry includes the cutoff-matter stress tensor to all orders in backreaction. This backreacting matter sector supplies a color variable distinct from the defect volume. We demonstrate the mechanism explicitly for the quantum BTZ black hole. Thus quantum backreaction resolves the same color-volume degeneracy addressed by the recent Weyl-factor proposal, but without introducing a non-standard boundary Weyl factor. Instead, the missing thermodynamic direction is supplied by the physical cutoff-matter sector of the doubly holographic quantum black hole. 
\end{abstract}

\maketitle

\textit{Introduction.---}A central problem in holographic quantum gravity is to identify which boundary variables faithfully encode bulk gravitational data.  The problem becomes especially delicate away from the strict classical limit, where bulk geometry is dressed by quantum fields and the gravitational entropy is replaced by a generalized entropy.  In recent years, quantum extremal surfaces, islands, and double holography have shown that semiclassical black holes can carry boundary information in ways that are invisible in a purely classical dictionary \cite{Ryu:2006bv,Engelhardt:2014gca,Penington:2019npb,Almheiri:2019hni}.  These developments also sharpen a more basic question: when a bulk parameter is varied, which boundary quantities represent genuinely independent physical directions, and which apparent variations are only artifacts of an incomplete dictionary?

Karch--Randall double holography is a particularly useful framework for this question \cite{Karch:2000ct,Karch:2000gx}.  A classical black hole in one higher dimension is reinterpreted as a lower-dimensional quantum black hole whose geometry includes the all-order backreaction of a cutoff CFT.  It therefore provides a controlled setting in which the thermodynamics of a black hole, the color of quantum matter, and the geometry of the system can be compared within a single dictionary.  We will use this fact to revisit one of the cleanest dictionary puzzles in black hole physics.

Extended black hole thermodynamics provides a setup in which to examine this question.  In this framework, the cosmological constant is treated as a pressure,
\begin{equation}
 P=-\frac{\Lambda}{8\pi G_N},
\end{equation}
and the gravitational first law contains a pressure-volume term \(V\dd P\) \cite{Kastor:2009wy,Kubiznak:2012wp,Kubiznak:2016qmn}. In AdS/CFT \cite{Maldacena:1997re,Witten:1998qj,Gubser:1998bc} this raises a sharp question: what is the boundary meaning of \(V\dd P\)?  The standard answer relates changes of the bulk AdS scale to changes of the number of boundary degrees of freedom, a viewpoint that also leads to a boundary description of black-hole phase transitions and criticality \cite{Visser:2021eqk,Cong:2021jgb}.  However, in an ordinary conformal frame this same change also changes the spatial size of the CFT.  For an Einstein bulk in \(d\) spacetime dimensions,
\begin{equation}
 C\sim \frac{L^{d-2}}{G_N}, \qquad \V_{d-2}\sim L^{d-2},
\end{equation}
when the boundary Weyl factor is fixed to \(\omega=1\).  Thus \(\dd C\) and \(\dd\V\) are not independent variations.  Ref.~\cite{Ahmed:2023snm} resolves this color-volume degeneracy by treating the boundary conformal factor as a thermodynamic variable, so that \(\V\sim(\omega L)^{d-2}\) (see Appendix~\ref{app:ordinary-degeneracy}).

\begin{figure}[t]
\centering
\begin{tikzpicture}[scale=0.92, every node/.style={font=\small}]
  \fill[blue!4] (-3.1,-1.45) rectangle (3.1,1.45);
  \draw[thick] (-3,-1.35) rectangle (3,1.35);
  \node at (0,1.15) {bulk \(\AdS_4\)};
  \draw[thick, blue!70!black] (-1.55,-1.25) .. controls (-0.95,-0.35) and (-0.95,0.35) .. (-1.55,1.25);
  \draw[thick, blue!70!black] (1.55,-1.25) .. controls (0.95,-0.35) and (0.95,0.35) .. (1.55,1.25);
  \node[blue!70!black] at (-1.65,-1.55) {brane \(B\)};
  \node[blue!70!black] at (1.65,-1.55) {brane \(B\)};
  \draw[thick, red!70!black] (-0.62,-0.62) circle (0.62);
  \draw[thick, red!70!black] (0.62,-0.62) circle (0.62);
  \node[red!70!black] at (0,-0.6) {horizon};
  \draw[dashed, thick] (-3,0) -- (-1.13,0);
  \draw[dashed, thick] (1.13,0) -- (3,0);
  \node at (-3.25,0) {\(\partial M\)};
  \node at (3.25,0) {\(\partial M\)};
  \node[align=center] at (0,1.75) {classical bulk black hole};
  \draw[-{Latex[length=2mm]}, thick] (0,1.55) -- (0,1.25);
  \node[align=center] at (0,-2.05) {quantum BTZ black hole on \(B\)\;\;\(\leftrightarrow\)\; defect \(\CFT_2\) + cutoff \(\CFT_3\)};
  \draw[-{Latex[length=2mm]}, thick] (0,-1.55) -- (0,-1.25);
\end{tikzpicture}
\caption{A cartoon showing Karch--Randall (KR) double holography.  Backreaction is built into the dictionary: a classical four-dimensional bulk solution with a brane is equivalently a semiclassical black hole on the brane, dressed by a cutoff matter sector.  The qBTZ black hole is the simplest analytic proof of concept for the resulting defect \(\CFT_2\) plus cutoff \(\CFT_3\) thermodynamics.}
\label{fig:double}
\end{figure}
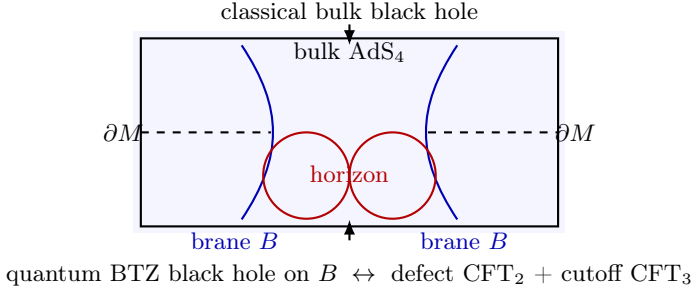

In this Letter, we show that the same degeneracy is avoided in Karch--Randall double holography for a more physical reason.  The starting point is the general structure of braneworld holography, not the special properties of a particular black hole.  Replacing the asymptotic regulator surface by a physical AdS brane integrates out the ultraviolet portion of the ambient CFT and induces a brane theory of the form
\begin{equation}
 I_{\rm eff}[B]=I_{\rm grav}[B]+I_{\rm CFT}[B].
 \label{eq:braneeff-main}
\end{equation}
Consequently, a classical bulk black hole ending on the brane is reinterpreted as a lower-dimensional quantum black hole whose geometry solves semiclassical equations with the cutoff CFT stress tensor included to all orders in backreaction \cite{Emparan:2002px,Emparan:2020znc,Frassino:2022zaz}.  In the doubly holographic description the same system is a defect \(\CFT_2\) coupled to a cutoff \(\CFT_3\).  This makes the relevant state space intrinsically two-scale: the brane/cutoff scale \(\ell\) controls the strength and color of the backreacting matter sector, while the brane curvature scale controls the size of the defect system (see Appendix~\ref{app:backreaction}).  The double-holographic setup is shown
schematically in Fig.~\ref{fig:double}. 

These two physical parameters allow the color and volume to vary independently even when the boundary metric is kept in the fixed representative $\omega=1$. Below we formulate this mechanism generally and then use the quantum BTZ black hole only as a solvable proof of concept (see Appendix~\ref{app:qbtz-scales}).

\textit{Backreaction as a braneworld feature.---}The induced description \eqref{eq:braneeff-main} shows why backreaction is natural in this setup. The cutoff CFT is not an external probe added by hand. It is part of the low-energy theory on the brane and arises naturally when the bulk region outside the brane is integrated out. Schematically, the brane equations take the semiclassical form
\begin{equation}
 E_{ab}^{\rm grav}[h]=8\pi G_{\rm br}\,\langle T_{ab}\rangle_{\rm cutoff},
 \label{eq:semiclassicalbrane}
\end{equation}
with higher-curvature terms in \(E_{ab}^{\rm grav}\) determined by the induced gravitational action. Therefore, the backreacting matter sector and the defect geometry provide two independent inputs to the holographic dictionary. We then ask whether the corresponding map to boundary thermodynamic variables has rank one or rank two. Here, rank one means that the resulting boundary variations are not independent, whereas rank two means that they span two independent thermodynamic directions.

\textit{qBTZ as a proof of concept.---}We now use the quantum BTZ solution obtained from an \(\AdS_4\) C-metric with an \(\AdS_3\) brane as the simplest analytic realization of this general mechanism \cite{Emparan:1999wa,Emparan:1999fd,Emparan:2020znc,Panella:2024sor}.  The brane black hole has line element
\begin{equation}
\begin{split}
 &\dd s_3^2=-f(r)\dd t^2+\frac{\dd r^2}{f(r)}+r^2\dd\phi^2,
 \\&
 f(r)=\frac{r^2}{\ellthree^2}-8G_3 M-\frac{\ell F(M)}{r},
 \label{eq:qbtzmetric}
 \end{split}
\end{equation}
where \(F(M)\) is the dimensionless qBTZ backreaction function
fixed by the parent C-metric and will not be needed below. \(\ell\) controls the quantum backreaction of the cutoff matter sector and \(\ellthree\) is the curvature scale appearing in the solution.  It is useful to introduce
\begin{equation}
 \nu\equiv \frac{\ell}{\ellthree} .
\end{equation}
We regard \((\ell,\ell_3)\) as local coordinates on the brane/cutoff parameter space at fixed renormalized \(G_3\) and verify whether their image in boundary thermodynamic
variables has rank one or rank two.

The color of the cutoff \(\CFT_3\), normalized as in Refs.~\cite{Emparan:2020znc,Frassino:2022zaz}, is
\begin{equation}
 c_3=\frac{L_4^2}{G_4}=\frac{\ell}{2G_3\sqrt{1+\nu^2}} .
 \label{eq:c3}
\end{equation}
This is the number of light degrees of freedom of the brane matter sector.  Here \(c_3\) denotes the stress-tensor normalization, or color,
of the cutoff CFT$_3$ sector; it is distinct from the
Brown--Henneaux central charge \(c_2=3L_3/(2G_3)\) of the
defect CFT$_2$.
The defect \(\CFT_2\) lives on a circle of radius \(L_3\).  We choose the fixed conformal representative
\begin{equation}
 \dd s^2_{\partial B}=-\dd t^2+L_3^2\dd\phi^2,
 \qquad \omega=1,
 \label{eq:omegaone}
\end{equation}
and hence
\begin{equation}
 \V_2=2\pi L_3 .
 \label{eq:volume}
\end{equation}
The radius \(L_3\) appearing in the brane action is not identical to \(\ellthree\); for weak backreaction,
\begin{equation}
 \frac{1}{L_3^2}=\frac{1}{\ellthree^2}\left[1+\frac{\ell^2}{4\ellthree^2}+O\left(\frac{\ell^4}{\ellthree^4}\right)\right].
 \label{eq:L3ell3}
\end{equation}
Thus both \(c_3\) and \(\V_2\) are functions of \(\ell\) and \(\ellthree\).  The crucial point is that they are not the same function (see Appendix~\ref{app:L3-ell3}).

\begin{figure}[t]
\centering
\begin{tikzpicture}[scale=0.96, every node/.style={font=\small}]
  \draw[->, thick] (-0.2,0) -- (3.2,0) node[right] {\(\log\ell\)};
  \draw[->, thick] (0,-0.2) -- (0,2.7) node[above] {\(\log\ell_3\)};
  \draw[->, very thick, blue!70!black] (0.7,0.55) -- (2.5,0.75) node[right] {mostly \(\dd\log c_3\)};
  \draw[->, very thick, red!70!black] (0.7,0.55) -- (0.95,2.25) node[above] {\qquad mostly \(\dd\log\mathcal V_2\)};
  \node[align=center] at (1.8,-0.55) {physical parameter-space};
  \draw[->, thick] (4.0,1.2) -- (4.8,1.2);
  \draw[->, thick] (5.4,0) -- (8.7,0) node[right] {\(\log c_3\)};
  \draw[->, thick] (5.6,-0.2) -- (5.6,2.7) node[above] {\(\log\mathcal V_2\)};
  \draw[->, very thick, blue!70!black] (6.05,0.55) -- (8.0,0.7);
  \draw[->, very thick, red!70!black] (6.05,0.55) -- (6.25,2.15);
  \node[align=center] at (7.25,-0.55) {boundary thermodynamics};
  \node at (4.4,1.55) {\(\J\neq0\)};
\end{tikzpicture}
\caption{The qBTZ map from physical parameter-space to boundary thermodynamic variables has nonzero rank.  The color direction \(c_3\) and the defect-volume direction \(\V_2\) are non-degenerate even though the boundary conformal factor is fixed.}
\label{fig:jacobian}
\end{figure}
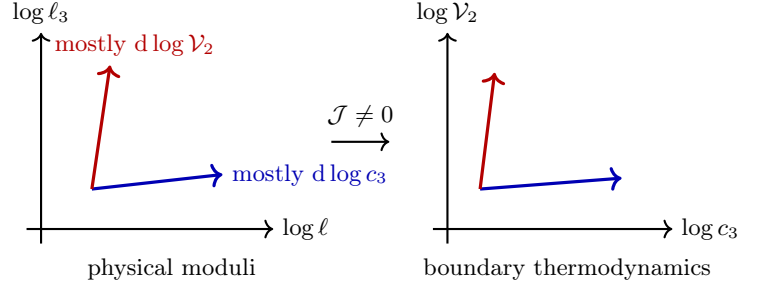

\textit{Nondegeneracy criterion.---}A genuine degeneracy occurs only if the map from physical parameters to boundary thermodynamic variables has rank one. Define
\begin{equation}
 \J\equiv \det\frac{\partial(\log c_3,\log\V_2)}{\partial(\log\ell,\log\ellthree)} .
 \label{eq:jacdef}
\end{equation}
Using Eqs.~\eqref{eq:c3}--\eqref{eq:L3ell3}, we find
\begin{align}
 \dd\log c_3
 &=\frac{1}{1+\nu^2}\dd\log\ell
 +\frac{\nu^2}{1+\nu^2}\dd\log\ellthree,
 \label{eq:varc3}\\
 \dd\log\V_2
 &=-\frac{\nu^2}{4}\dd\log\ell
 +\left(1+\frac{\nu^2}{4}\right)\dd\log\ellthree
 +O(\nu^4).
 \label{eq:varV2}
\end{align}
Therefore
\begin{equation}
 \J=1-\frac{3}{4}\nu^2+O(\nu^4)\neq 0
 \label{eq:jacobian}
\end{equation}
for small finite backreaction.  The corresponding rank-two map from physical parameters to boundary
thermodynamic variables is illustrated in Fig.~\ref{fig:jacobian}. The two variations are linearly independent.  In particular, varying \(\ell\) changes the cutoff color at leading order while leaving the defect volume fixed up to \(O(\nu^2)\), whereas varying \(\ellthree\) changes the defect volume at leading order.  This is the desired nondegenerate structure, obtained without varying a boundary Weyl factor (see Appendix~\ref{app:L3-ell3}).

\textit{First law.---}The higher-dimensional origin of the brane pressure is the tension work term.  Varying the brane tension changes the effective cosmological constant on the brane and induces the extended qBTZ first law
\begin{equation}
 \dd M=T\dd\Sgen+V_3\dd P_3
 \label{eq:brane-fixed-c3}
\end{equation}
in the fixed-\(c_3\) ensemble \cite{Frassino:2022zaz}.  Allowing the cutoff/color parameter to vary defines the enlarged
double-holographic ensemble. In this ensemble \(M\) is treated as a
thermodynamic potential on the local state space
\((S_{\rm gen},P_3,c_3)\), so that
\[
dM=T dS_{\rm gen}+V_3 dP_3+\mu_3 dc_3,\qquad
\mu_3=\left(\frac{\partial M}{\partial c_3}\right)_{S_{\rm gen},P_3}.
\]

The corresponding three-dimensional Smarr relation has no mass term because \(G_3M\) is dimensionless (see Appendix~\ref{app:first-laws}),
\begin{equation}
 0=T\Sgen-2P_3V_3+\mu_3c_3 .
 \label{eq:qbtz-smarr}
\end{equation}
The doubly holographic description rewrites the same thermodynamics as that of a defect \(\CFT_2\) on the circle \(\V_2=2\pi L_3\), coupled to a cutoff \(\CFT_3\) with color \(n\equiv c_3\).  The most general boundary first law in the sector considered here takes the schematic form
\begin{equation}
 \dd E=T\dd S_2-p_2\dd\V_2+\mu_2\dd c_2+\mu_n\dd n .
 \label{eq:cft2-law}
\end{equation}
At fixed \(G_3\), the defect Brown--Henneaux central charge \(c_2=3L_3/(2G_3)\) is proportional to \(\V_2\).  Hence the pair \((c_2,\V_2)\) is indeed degenerate.  This is not the relevant obstruction.  The quantum black hole has a second color variable, \(n=c_3\), associated with the light cutoff matter sector.  Equations~\eqref{eq:varc3}--\eqref{eq:jacobian} show that \((n,\V_2)\), not \((c_2,\V_2)\), forms a nondegenerate thermodynamic pair (see Appendix~\ref{app:first-laws}).

\textit{Physical interpretation.---}In ordinary boundary CFT thermodynamics, introducing an independent \(\omega\) separates ``how many sites'' from ``how many degrees of freedom per site.''  In the Karch--Randall quantum black hole, this separation is not imposed by a choice of conformal representative; it is generated by the semiclassical state itself.  The scale \(\ell\) changes the UV cutoff and the number of light degrees of freedom whose stress tensor backreacts on the brane.  The brane curvature scale controls the size of the defect system.  The distinction between \(c_3\) and \(\V_2\) is intrinsically semiclassical: \(c_3\) characterizes the backreacting cutoff matter sector, while \(\V_2\) measures the spatial volume of the lower-dimensional defect theory.  The two deformations are physical, corresponding to moving the brane/cutoff and changing the induced worldvolume curvature, not to choosing a different representative of the same boundary conformal class.

This distinction is especially transparent in the small-backreaction regime \(\nu\ll1\):
\begin{equation}
 \dd\log c_3=\dd\log\ell+O(\nu^2),
 \qquad
 \dd\log\V_2=\dd\log\ellthree+O(\nu^2).
 \label{eq:leading}
\end{equation}
Therefore, even with \(\omega=1\), one can vary the cutoff color at fixed defect volume or vary the defect volume at fixed cutoff color.  Black hole chemistry is then nondegenerate for a physical reason intrinsic to double holography.

\textit{Discussion.---}We have argued that quantum black holes on Karch--Randall branes naturally avoid the color-volume degeneracy of ordinary holographic black hole chemistry.  The essential ingredient is not a boundary Weyl factor, but the fact that braneworld holography automatically produces a semiclassical brane black hole dressed by a finite, backreacting cutoff matter sector.  The qBTZ calculation is a proof of concept for this general statement: in the simplest analytic example the physical parameters \((\ell,\ell_3)\) map with nonzero Jacobian to the boundary variables \((c_3,\V_2)\).  The result suggests that holographic black hole chemistry is naturally enlarged in the semiclassical theory: the pressure-volume sector describes the geometry of the defect system, while the color-chemical-potential sector describes the quantum matter responsible for backreaction.

The same viewpoint is timely for charged and rotating quantum black holes.  Recent work has developed the thermodynamics and chemical potentials of charged braneworld quantum black holes \cite{Feng:2024uia,Climent:2024nuj} and has extended these constructions to charged rotating solutions \cite{Bhattacharya:2025tdn}. Charged quantum black holes have also been studied in de Sitter braneworlds, including their nucleation and thermodynamics \cite{Climent:2024wol}.  Independently, the phase structure and thermodynamic geometry of qBTZ black holes have revealed critical behavior that differs from standard mean-field Van der Waals universality \cite{HosseiniMansoori:2024bfi}. Our parameter-space criterion identifies which variables remain independent when the phase diagram is extended to charged or rotating black holes. In particular, pressure, defect volume, cutoff-CFT color, and charge should not be tied to a single length scale. This opens a systematic route to quantum black hole phase transitions in which the ensemble is specified directly in terms of physical brane/cutoff parameters (see Appendix~\ref{app:charged-critical}).  The mechanism should also extend to higher-dimensional brane black holes and DGP-deformed branes \cite{Dvali:2000hr} where the separation between gravitational strength, graviton mass, and cutoff color is even richer. Further details on the parameter-space calculation, the enlarged first law, and the comparison with Weyl-factor thermodynamics are given in Appendices~\ref{app:L3-ell3}, \ref{app:first-laws}, and \ref{app:weyl-factor}.

\appendix
\section{Ordinary AdS/CFT degeneracy at fixed Weyl frame}
\label{app:ordinary-degeneracy}

For an ordinary holographic CFT dual to Einstein gravity in \(d\) bulk spacetime dimensions, the central charge and spatial volume scale as \cite{Karch:2015rpa,Visser:2021eqk,Ahmed:2023snm}
\begin{equation}
 C\sim \frac{L^{d-2}}{G_N},
 \qquad
 \V\sim (\omega L)^{d-2}.
 \label{eq:suppordinary}
\end{equation}
At fixed \(G_N\) and fixed \(\omega=1\),
\begin{equation}
 \dd\log C=(d-2)\dd\log L,
 \qquad
 \dd\log\V=(d-2)\dd\log L.
\end{equation}
Thus \(\dd C/C=\dd\V/\V\). A first law containing both \(\mu\dd C\) and \(-p\dd\V\) is therefore kinematically degenerate unless a second independent parameter is introduced. Ref.~\cite{Ahmed:2023snm} uses the boundary Weyl factor \(\omega\) as this additional parameter. The present Letter instead uses the two independent physical parameters already available in Karch--Randall double holography.

\section{Backreaction as a generic braneworld output}
\label{app:backreaction}

The argument in the main text does not assume the qBTZ metric at the outset.  In braneworld holography, the brane replaces a radial regulator surface and makes the cutoff physical.  Integrating out the exterior region induces a gravitational action on the brane together with a cutoff CFT sector \cite{Emparan:2002px,Emparan:2020znc,Chen:2020uac},
\begin{equation}
 I_{\rm eff}[B]=I_{\rm grav}[B]+I_{\rm CFT}[B].
 \label{eq:endbraneeff}
\end{equation}
Because the brane metric is dynamical, the cutoff CFT stress tensor contributes directly to the gravitational equations rather than behaving as an external source. Thus a smooth classical solution of the higher-dimensional bulk-plus-brane problem gives a lower-dimensional black hole that already includes quantum matter backreaction \cite{Emparan:2002px,Emparan:2020znc}.  This is the sense in which the brane black hole is a quantum black hole.

The thermodynamic implication is that the matter color and the defect volume are not forced to be described by a single AdS radius.  The cutoff/brane position controls the number and coupling strength of light cutoff degrees of freedom, while the induced worldvolume curvature controls the size of the defect theory.  A one-scale degeneracy can still occur in special subsectors, for example between \(c_2\) and \(\V_2\) at fixed \(G_3\), but it is not a generic feature of the full double-holographic state space.  The qBTZ solution provides the cleanest analytic example in which this statement can be checked explicitly.

\section{qBTZ as a proof of concept: scales and central charges}
\label{app:qbtz-scales}

The quantum BTZ black hole is obtained by cutting off an \(\AdS_4\) C-metric with an \(\AdS_3\) Karch--Randall brane \cite{Emparan:1999wa,Emparan:1999fd,Emparan:2020znc,Panella:2024sor}.  The induced brane geometry is
\begin{equation}
\begin{split}
& \dd s_3^2=-f(r)\dd t^2+\frac{\dd r^2}{f(r)}+r^2\dd\phi^2,
 \\&
 f(r)=\frac{r^2}{\ell_3^2}-8G_3M-\frac{\ell F(M)}{r}.
 \end{split}
\end{equation}
The parameter \(\ell\) measures the position of the brane/cutoff and controls quantum backreaction; \(\ell_3\) is the curvature scale appearing in the qBTZ solution.  They combine into
\begin{equation}
 \nu=\frac{\ell}{\ell_3}.
\end{equation}
The color of the cutoff \(\CFT_3\) sector is \cite{Emparan:2020znc,Frassino:2022zaz}
\begin{equation}
 c_3=\frac{L_4^2}{G_4}=\frac{\ell}{2G_3\sqrt{1+\nu^2}}.
\end{equation}
The defect central charge is instead \cite{1986CMaPh.104..207B}
\begin{equation}
 c_2=\frac{3L_3}{2G_3},
 \label{eq:c2supp}
\end{equation}
and the defect spatial volume in the fixed conformal frame is
\begin{equation}
 \V_2=2\pi L_3.
 \label{eq:V2supp}
\end{equation}
The distinction between \(c_2\) and \(c_3\) is essential.  The former is the Virasoro central charge of the defect theory; the latter counts the cutoff matter degrees of freedom that backreact on the brane geometry.  The black hole is quantum because this second sector is dynamical and finite.

\section{Relation between \(L_3\) and \(\ell_3\)}
\label{app:L3-ell3}

The radius \(L_3\) appearing in the brane action differs from the solution radius \(\ell_3\) in higher-curvature induced gravity \cite{Emparan:2020znc,Frassino:2022zaz}.  In the weak-backreaction expansion,
\begin{equation}
 \frac{1}{L_3^2}=\frac{1}{\ell_3^2}
 \left(1+\frac{\ell^2}{4\ell_3^2}+O\left(\frac{\ell^4}{\ell_3^4}\right)\right).
\end{equation}
Consequently,
\begin{equation}
 L_3=\ell_3\left(1-\frac{\nu^2}{8}+O(\nu^4)\right),
 \qquad
 \V_2=2\pi\ell_3\left(1-\frac{\nu^2}{8}+O(\nu^4)\right).
\end{equation}
Therefore the defect volume does depend on both \(\ell\) and \(\ell_3\) beyond leading order.  Nondegeneracy does not require \(\V_2\) to depend on \(\ell_3\) alone; it requires the dependence of \(\V_2\) and \(c_3\) on \((\ell,\ell_3)\) to be linearly independent.

\subsection*{Variation formulae}

At fixed \(G_3\),
\begin{equation}
 \log c_3=\log\ell-\frac{1}{2}\log(1+\nu^2)+\mathrm{const}.
\end{equation}
Since \(\dd\log\nu=\dd\log\ell-\dd\log\ell_3\),
\begin{align}
 \dd\log c_3
 &=\dd\log\ell-\frac{\nu^2}{1+\nu^2}\dd\log\nu \\
 &=\frac{1}{1+\nu^2}\dd\log\ell
 +\frac{\nu^2}{1+\nu^2}\dd\log\ell_3.
\end{align}
For the volume,
\begin{equation}
 \log\V_2=\log\ell_3-\frac{\nu^2}{8}+O(\nu^4)+\mathrm{const},
\end{equation}
and hence
\begin{align}
 \dd\log\V_2
 &=\dd\log\ell_3-\frac{\nu^2}{4}\dd\log\nu+O(\nu^4)\\
 &=-\frac{\nu^2}{4}\dd\log\ell
 +\left(1+\frac{\nu^2}{4}\right)\dd\log\ell_3+O(\nu^4).
\end{align}
The Jacobian is therefore
\begin{align}
 \J
 &=
 \det
 \begin{pmatrix}
 \dfrac{1}{1+\nu^2} & \dfrac{\nu^2}{1+\nu^2} \\
 -\dfrac{\nu^2}{4} & 1+\dfrac{\nu^2}{4}
 \end{pmatrix}
 +O(\nu^4) \\
 &=1-\frac{3}{4}\nu^2+O(\nu^4).
\end{align}
Thus \(\J\neq0\) in the perturbative quantum regime.

A slightly more general way to see this is to write
\begin{equation}
 c_3=\frac{\ell}{G_3}f(\nu),
 \qquad
 \V_2=\ell_3 h(\nu).
\end{equation}
Then
\begin{equation}
 \J=1+\frac{\dd\log f}{\dd\log\nu}-\frac{\dd\log h}{\dd\log\nu}.
\end{equation}
Thus degeneracy requires $\J=0$ and is therefore a codimension-one condition on the functions \(f\) and \(h\), rather than a generic consequence of both quantities depending on both scales.

\section{Brane and boundary first laws}
\label{app:first-laws}

With fixed bulk pressure and variable brane tension, the bulk first law contains a brane work term \cite{Frassino:2022zaz},
\begin{equation}
 \dd M=T\dd S+A_\tau\dd\tau.
\end{equation}
 Because the brane tension determines the effective cosmological constant on the brane, its variation leads to the extended qBTZ first law \cite{Frassino:2022zaz}
\begin{equation}
\dd M=T\dd\Sgen+V_3\dd P_3
\end{equation}
in the fixed-$c_3$ ensemble. Allowing the cutoff color to vary gives
\begin{equation}
 \dd M=T\dd\Sgen+V_3\dd P_3+\mu_3\dd c_3.
\end{equation}
The conjugates are defined by
\begin{equation}
 V_3=\left(\frac{\partial M}{\partial P_3}\right)_{\Sgen,c_3},
 \qquad
 \mu_3=\left(\frac{\partial M}{\partial c_3}\right)_{\Sgen,P_3}.
\end{equation}
The corresponding Smarr law is \cite{Frassino:2022zaz}
\begin{equation}
 0=T\Sgen-2P_3V_3+\mu_3c_3.
\end{equation}
The absence of a mass term is characteristic of three-dimensional black hole thermodynamics because \(G_3M\) is dimensionless.

The boundary defect theory has \cite{1986CMaPh.104..207B,Emparan:2020znc}
\begin{equation}
 c_2=\frac{3L_3}{2G_3},
 \qquad
 \V_2=2\pi L_3.
\end{equation}
Thus, if only \(c_2\) and \(\V_2\) are retained at fixed \(G_3\), their variations are degenerate:
\begin{equation}
 \frac{\dd c_2}{c_2}=\frac{\dd\V_2}{\V_2}.
\end{equation}
The doubly holographic theory, however, also contains the cutoff color \(n\equiv c_3\).  The nondegenerate boundary first law is therefore naturally organized as
\begin{equation}
 \dd E=T\dd S_2-p_2\dd\V_2+\tilde\mu_2\dd c_2+\tilde\mu_n\dd n.
\end{equation}
Equivalently, since \(c_2\propto\V_2\), the first two work terms may be combined into an effective pressure term, while \(\tilde\mu_n\dd n\) remains independent.  This is the operational sense in which quantum black holes on the brane avoid the ordinary color-volume degeneracy.

\section{Extension to charged and critical quantum black holes}
\label{app:charged-critical}

The nondegeneracy criterion is especially useful once additional thermodynamic sectors are present.  Charged quantum black holes constructed from the AdS charged C-metric introduce electric and magnetic parameters in addition to the brane/cutoff scales.  Their doubly holographic first laws contain the usual charge-potential pairs together with the color variables associated with the defect and cutoff matter sectors \cite{Feng:2024uia,Climent:2024nuj}.  Rotating charged solutions further enlarge this parameter space \cite{Bhattacharya:2025tdn}.  In such a setting, a one-scale boundary dictionary can obscure which quantities are independently varied in a canonical, grand-canonical, or mixed ensemble.  The parameter-space formulation outlined here instead begins with the physical parameters of the bulk/brane construction and then computes the rank of the map to boundary thermodynamic variables.

This is also relevant for phase transitions.  Existing qBTZ studies have already found nonstandard critical behavior in extended thermodynamics \cite{HosseiniMansoori:2024bfi}, and charged quantum black holes provide a natural arena in which quantum backreaction, charge, and pressure compete \cite{Feng:2024uia}.  The present framework suggests that the appropriate phase space should distinguish the defect volume \(\V_2\), the cutoff color \(c_3\), the defect central charge \(c_2\), and any charge sector before imposing ensemble constraints.  Degeneracies should then be derived from the Jacobian of this map, rather than assumed from the ordinary single-radius AdS/CFT dictionary.

\section{Relation to Weyl-factor thermodynamics}
\label{app:weyl-factor}

The Weyl-factor proposal \cite{Ahmed:2023snm} and the present mechanism both enlarge the thermodynamic state space from one parameter to two. The difference lies in the origin of the second parameter:
\begin{subequations}\label{app:weyl-maps}
\begin{align}
\text{ordinary AdS/CFT:}\qquad
&(L,\omega)\mapsto (C,\V), \label{eq:G1}\\
\text{KR double holography:}\qquad
&(\ell,\ell_3)\mapsto (c_3,\V_2). \label{eq:G2}
\end{align}
\end{subequations}

In the first case, the additional parameter is the boundary conformal factor. In the second case, both parameters are physical bulk/brane quantities associated with the cutoff brane and the induced worldvolume geometry.  Therefore braneworld quantum black holes can possess a nondegenerate holographic first law while keeping the boundary metric in the fixed representative \(\omega=1\). The qBTZ construction is the explicit proof of concept in which the rank-two map can be constructed analytically.

\bibliographystyle{utphys1}
\bibliography{bib}

\end{document}